\documentclass{article}
\usepackage{spconf,amsmath,graphicx}
\usepackage{subfig}
\usepackage{graphicx}
\usepackage{url}

\title{mmWave-Whisper: Phone Call Eavesdropping and Transcription \\Using Millimeter-Wave Radar}
\name{Suryoday Basak, Abhijeeth Padarthi, Mahanth Gowda}
\address{Pennsylvania State University, University Park, PA, USA}

\usepackage{xcolor}

\newcommand{\papertitle}{\emph{mmWave-Whisper}}

\begin{document}
%
\maketitle
\begin{abstract}

This paper introduces {\papertitle}, a system that demonstrates the feasibility of full-corpus automated speech recognition (ASR) on phone calls eavesdropped remotely using off-the-shelf frequency modulated continuous wave (FMCW) millimeter-wave radars. Operating in the 77-81 GHz range, {\papertitle} captures earpiece vibrations from smartphones, converts them into audio, and processes the audio to produce speech transcriptions automatically. Unlike previous work that focused on loudspeakers or limited vocabulary, this is the first work to perform such a speech recognition by handling large vocabulary and full sentences on earpiece vibrations from smartphones. This approach expands the potential of radar-audio eavesdropping. {\papertitle} addresses challenges such as the lack of large scale training datasets, low SNR, and limited frequency information in radar data through a systematic pipeline designed to leverage synthetic training data, domain adaptation, and inference by incorporating OpenAI's Whisper automatic speech recognition model. The system achieves a word accuracy rate of 44.74\% and a character accuracy rate of 62.52\% over a range of 25 cm to 125 cm. The paper highlights emerging misuse modalities of AI as the technology evolves rapidly. 

\end{abstract}
\begin{keywords}
side channel attacks, mmWave radars, speech privacy, speech recognition
\end{keywords}

\section{Introduction}\label{sec:intro}

Millimeter-wave (mmWave) technology is crucial for next-generation applications requiring high bandwidth, such as 5G/6G networks, AR/VR, ADAS, and health monitoring. As it becomes more common in commercial systems like smartphones, electric vehicles, and smart devices, this paper addresses a significant concern: the potential misuse of mmWave radars for eavesdropping on phone calls, revealing sensitive information.


This paper presents {\papertitle}, a system that leverages off-the-shelf mmWave radar devices to eavesdrop on audio spoken by remote callers during phone conversations and transcribes the captured sentences through an automatic speech recognition (ASR) model. By detecting minute vibrations ($\approx 7 \mu m$) generated by the phone’s earpiece, converting them into audio, and then transcribing them, {\papertitle} captures audio content that is otherwise inaudible to both humans and traditional microphones. Its immunity to ambient noise makes it particularly effective in noisy environments, raising significant security concerns as phone calls, typically considered secure, can be intercepted and confidential information compromised. 

\noindent
\textbf{Challenges:} The implementation of {\papertitle} involves addressing several unique challenges. The earpiece vibrations detected by mmWave radars are extremely small, lying close to the radar’s noise floor, making them difficult to isolate. Additionally, noise from frequency oscillators and multipath signal interference complicate clean audio extraction. The lack of large-scale radar-audio datasets further limits model training, and the extracted audio is heavily degraded, with an SNR less than 5 dB and a frequency range below 1.5kHz -- far below what conventional ASR models require. These combined challenges make this a distinctly complex problem to address.

\noindent

\noindent
\textbf{Solutions:} To tackle these challenges, {\papertitle} implements several novel solutions through a carefully designed system. By tracking stable reflections over time, we isolate the phone’s weak vibrations from surrounding noise, and apply statistical error correction to reduce interference from frequency oscillators. By exploiting phase variations in the radar signal, we capture micrometer-level changes, allowing the conversion of subtle vibrations into usable audio streams. In response to the lack of large-scale radar-audio datasets, we generate synthetic radar-audio data and apply domain adaptation to the ASR model, further refining it with a small sample of real radar data. Despite the audio’s limited frequency range, the system effectively adapts to these constraints, demonstrating that accurate speech recognition is achievable even under extreme conditions.

Prior works on speech sensing and extraction using radio waves include RadioMic \cite{radiomic}, WaveEar \cite{waveear}, and Wireless Vibrometry \cite{wireless_vib}. mmPhone \cite{mmphone} focuses on acoustic eavesdropping on loudspeakers using an mmWave radar and a piezoelectric film. WavoID \cite{wavoid}, Wavoice \cite{wavoice}, and RadioSES \cite{radioses} demonstrate multimodal methods combining microphones and radars for user identification, speech recognition, and source separation, respectively. Wavesdropper \cite{wavesdropper} achieves ASR with human subjects in a through-wall setting, while Radio2Text \cite{radio2text} introduces a knowledge distillation method for large-corpus speech recognition from loudspeakers. mSilent \cite{mSilent} presents a multimodal system combining video and mmWave radar for silent speech recognition. Although mmSpy \cite{mmspy} and mmEve \cite{mmeve} expose similar phone call eavesdropping attacks, they differ significantly from {\papertitle} in their scope and approach. mmSpy focuses on individual keywords and digits, while mmEve uses 100 Harvard Sentences, with overlapping utterances between the training and test sets, limiting its generalizability to users not in the dataset. In contrast, {\papertitle} employs data from LibriSpeech \cite{librispeech}, which features a full 10,000-word corpus and sentences of varying lengths. To ensure our system is more generalizable to unseen audio and speakers, we enforce complete non-overlap between the utterances and speakers in the training and test data. With this configuration, {\papertitle} achieves a word accuracy rate of 44.74\% and a character accuracy rate of 62.52\%, demonstrating its ability to handle previously unseen data effectively.


Since this research addresses security concerns, even partial word matches pose a privacy risk. While ASR systems typically aim for high accuracy, the extraction of any intelligible information from radar-audio is a significant security threat, making data leakage the primary focus over maximal recognition accuracy.


In achieving the reported levels of attack accuracy on smartphones using off-the-shelf radar devices, we highlight our key contributions as follows: \begin{itemize} 
  \itemsep0em
  \item To the best of our knowledge, this is the first work to explore security threats related to eavesdropping on phone call earpiece devices using mmWave radars with full corpus speech data.
  \item We implemented a system specifically designed and domain-adapted for radar-audio speech inference, that incorporates OpenAI’s Whisper ASR model, overcoming challenges such as the lack of open-source datasets, low SNR, and limited frequency range in the data.
  \item As a broader contribution, we show the feasibility of using ASR models on extremely noisy audio data.
\end{itemize}

\begin{figure}
    \centering
    \includegraphics[width=\columnwidth]{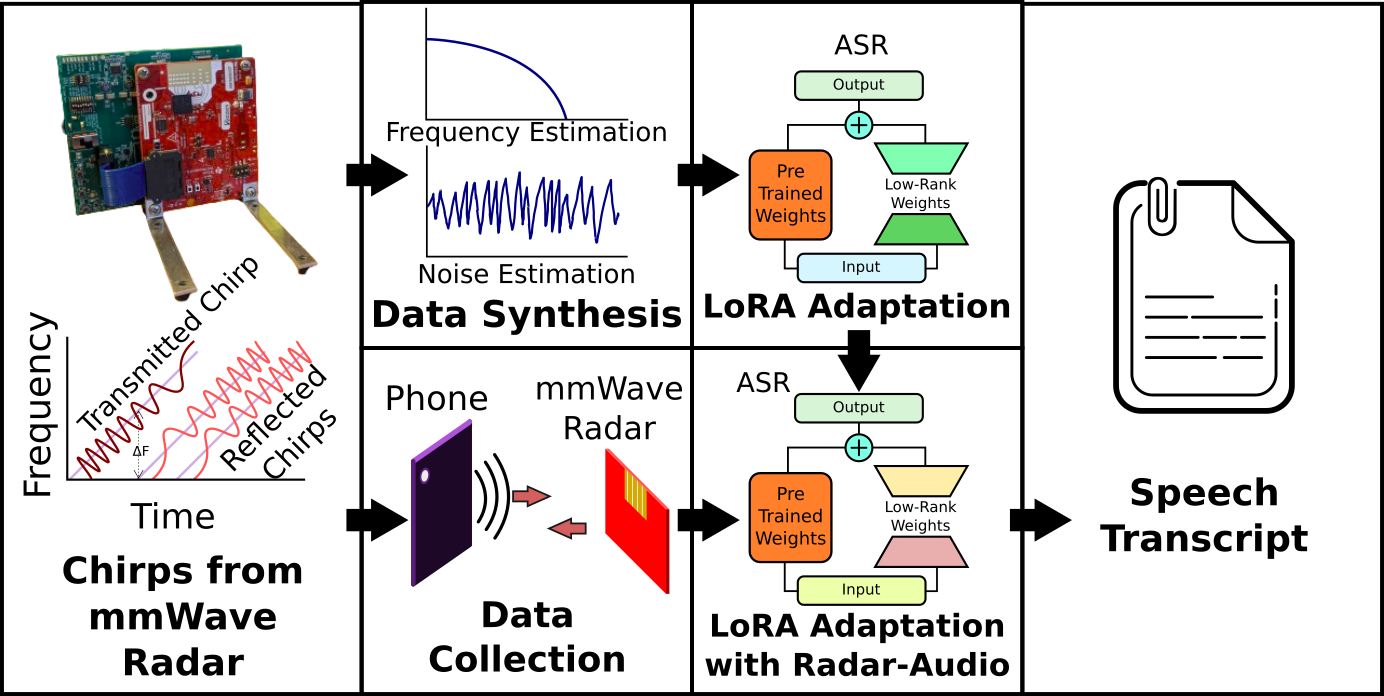}
    \caption{System design of {\papertitle}.}
    \label{fig:enter-label}
    \vspace{-0.6cm}
\end{figure}
\section{Background}\label{sec:background}

An FMCW radar emits signals with variable frequency components, called \textit{chirps}. Chirps reflect off nearby objects and are received by the radar after a time delay. The transmitted and received signals are mixed to produce an intermediate frequency (IF) signal, which is analyzed using the fast Fourier transform (FFT) to estimate object position and velocity from frequency components and Doppler shifts. When the FFT is used to estimate distance or position, it is called \textit{range-FFT} \cite{ti-mmwave}. Micrometer-level changes can be detected from the phase term in the IF signal, which can be sequenced over time to generate an audio stream.

Representing the distance between the phone and radar as $d$, the rate of chirp frequency increase as $S$, the wavelength of the chirp as $\lambda$, the amplitude of the IF signal as $A$, and the speed of light as $c$, the IF signal produced by an object at distance $d$ can be expressed as \cite{ti-mmwave}:

\begin{equation}
    IF(d) = A \sin (2 \pi f_0 t + \phi)
\end{equation}

where $f_0 = \frac{2Sd}{c}$ and $\phi=\frac{4\pi d}{\lambda}$. The vibrations of the phone are accounted as tiny changes in $d$. While the range-FFT enables extraction of $f_0$ and estimation of $d$, micrometer-level changes in $d$ are detected through $\phi$ as it can be measured at a high-precision (micrometers) which allows us to measure changes in $d$ (smartphone vibrations in micrometer granularity).

Smartphone earpiece vibrations, though much weaker than loudspeaker vibrations, propagate across the phone’s body due to compact design. {\papertitle} detects these subtle vibrations by extracting the phases from the mmWave radar signals reflected from the back of the phone, enabling eavesdropping even when the vibrations are not directly aligned with the radar. The earpiece vibrations detected by the radar have a very low signal-to-noise ratio (SNR), typically less than 5 dB. Moreover, the frequency range in the radar-audio data is clipped around 1.5kHz, which is below the frequency range expected by traditional ASR systems. To account for the frequency range observed in radar-audio, we configured the radar system with a 4kHz audio sampling rate, following a the methodology as outlined in \cite{mmspy}.

\section{Core Technical Modules}\label{sec:tech_modules}







\subsection{Audio Extraction and Data Collection}

Audio extraction from radar data primarily involves tracking phase variations in the IF signal generated by mmWave radars. The process starts by identifying and isolating the strongest peak in the range-FFT, which corresponds to the reflection from the phone’s body. This peak is then sequenced over time from successive radar frames to generate an audio stream, capturing subtle vibrations that represent the earpiece audio.

However, the phase data is prone to errors, particularly from power spikes at the start of each radar frame, which distort the vibration signals. To address this, statistical error correction \cite{mmspy} is applied to reduce phase fluctuations and correct for power spikes and DC drifts, enabling clearer audio extraction. By combining phase variation tracking, peak isolation, and error correction, the system effectively reduces noise caused by the nuances of mmWave radar design.



\subsection{Synthetic Data Generation}

To generate synthetic radar-audio data, we first filter the LibriSpeech audio using a Butterworth low-pass filter at 1.1kHz, based on our observation that frequency components start to decline sharply beyond this point. Next, we generate a random noise vector by estimating the noise amplitude from the radar-audio data, then add this noise vector to the filtered LibriSpeech audio. After this, we apply another low-pass filter at 2kHz, matching the Nyquist frequency of our radar’s audio sampling configuration. Since the Whisper ASR model expects input data at 16kHz, we upsample the filtered and noise-added audio to 16kHz. This process simulates radar-audio characteristics, allowing us to create a large-scale dataset for domain adaptation of the Whisper model, overcoming the absence of large-scale radar-audio datasets.

\subsection{Domain Adaptation of Whisper ASR Model}{\label{subsec:finetuning}}

Whisper models \cite{radford-whisper}, developed by OpenAI, are ASR systems for transcribing spoken language across various languages. Whisper-large is the largest-parameter version of the Whisper family of models 1.5B parameters. In {\papertitle}, we adapted the Whisper-large version 2 (Whisper-large-v2) model with synthetic data using LoRA, and did further training of the LoRA parameters with a small sample of real mmWave radar-audio data. The key steps in adapting Whisper-large for radar-audio are as follows:\\

\begin{figure}
    \centering
    \includegraphics[width=0.8\columnwidth]{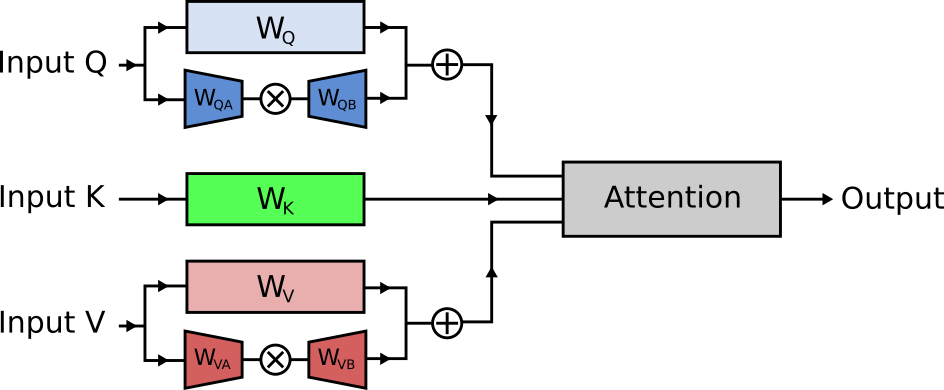}
    \caption{The attention blocks in {\papertitle} with LoRA parameters. $W_Q$, $W_K$, and $W_V$ represent the query, key, and value weight tensors, respectively. The low-rank tensors $W_{QA}$ and $W_{QB}$ correspond to the query tensor, while $W_{VA}$ and $W_{VB}$ correspond to the value tensor. LoRA is applied to both the self-attention and cross-attention modules.}
    \label{fig:lora-attention}
    \vspace{-0.5cm}
\end{figure}

\noindent
\textbf{(a) Low-Rank Adaptation (LoRA):} LoRA \cite{lora}, a method for efficiently fine-tuning large models, uses low-rank decomposition to reduce memory and computational requirements. It introduces two low-rank matrices, $W_A$ and $W_B$, which are injected in parallel to a weight tensor $W$ during training. The key idea is that the large parameter tensor $W$ is kept frozen, and only $W_A$ and $W_B$ are trained, significantly reducing the number of trainable parameters. In {\papertitle}, we apply LoRA to the value ($V$) and query ($Q$) tensors of attention blocks in the Whisper-large model, with the rank of the LoRA modules set to 32 and the scaling factor $\alpha$ set to 64. This results in only $1\%$ of the model's parameters being trained, allowing Whisper-large to fit within GPU memory while maintaining strong performance during domain adaptation. Figure \ref{fig:lora-attention} illustrates the attention modules in {\papertitle} after the application of LoRA.

\noindent
\textbf{(b) Training LoRA Parameters:} The LoRA parameters of Whisper-large-v2 was trained for 45 hours using 460 hours of synthetic data from the LibriSpeech clean-100 and clean-360 sets, with spectrogram data input containing 80 mel frequency bins. The training procedure starts with domain adapting the Whisper-large-v2 model using the synthetic radar-audio data to help the model adapt to the noisy and low-frequency characteristics of radar-captured audio. This initial step allows the model to learn the general patterns of radar audio while minimizing computational costs through the use of LoRA. After this synthetic-data training, the LoRA parameters were further trained with a small set of 772 real mmWave radar-audio utterances, ensuring the model captured the specific nuances of real-world radar-audio.

\subsection{Experimental Platform}
{\papertitle} is implemented using the Texas Instruments AWR1843 BOOST (77-81 GHz, FMCW signals) and tested on a Samsung Galaxy S20. Radar data is pre-processed with MATLAB and Python, then fed into ASR models built with HuggingFace Transformers and PyTorch. The LoRA parameters of Whisper-large-v2 are trained on an NVIDIA Quadro RTX 8000 GPU.
\section{Evaluation}\label{sec:eval}

\begin{table*}[t]
\resizebox{\linewidth}{!}{
    \centering
    \begin{tabular}{|c|c|c|c|}
    \hline
         \textbf{Generated Transcription} & \textbf{True Transcription} & $W_{acc}$ & $C_{acc}$\\
         \hline
         after that it was easy to forget actually to forget & after that it was easy to forget actually to forget & 100.0 & 100.0 \\
         \hline
         you haven t seen any of them no she & you haven t seen any of them no sir & 88.89 & 94.29 \\
         \hline
         that he will teach me a few tricks when water is coming & yet he will teach you a few tricks when morning is come & 66.67 & 70.91 \\
         \hline
         at the same time every avenue of the home was felt in his voice & at the same time every avenue of the throne was assaulted with gold & 61.54 & 73.13 \\
         \hline
        he had broken into the straight line & he had broken into her courtyard & 50.0 & 56.25 \\
         \hline
         oh sir what s different observed joseph arching and beneath it & oh say that s different observed markham altering his demeanor & 30.0 & 61.29 \\
         \hline
         her consciousness of her was now in order & his constancy to her was now rewarded & 28.57 & 54.05 \\
         \hline
         you ve got a bit of what wasn t his kind doing in there mister & you ve got a birthday present this time jim and no mistake & 8.33 & 46.55 \\
         \hline
    \end{tabular}
    }
    \caption{Sample transcriptions generated by {\papertitle} compared to the original sentences from the LibriSpeech dataset, illustrating the system's performance in noisy radar-audio conditions, along with their word accuracy rate ($W_{acc}$) and character accuracy rate ($C_{acc}$).}
    \label{tab:utterances_table}
    \vspace{-0.4cm}
\end{table*}

\begin{figure}
    \centering
    \includegraphics[width=0.8\columnwidth]{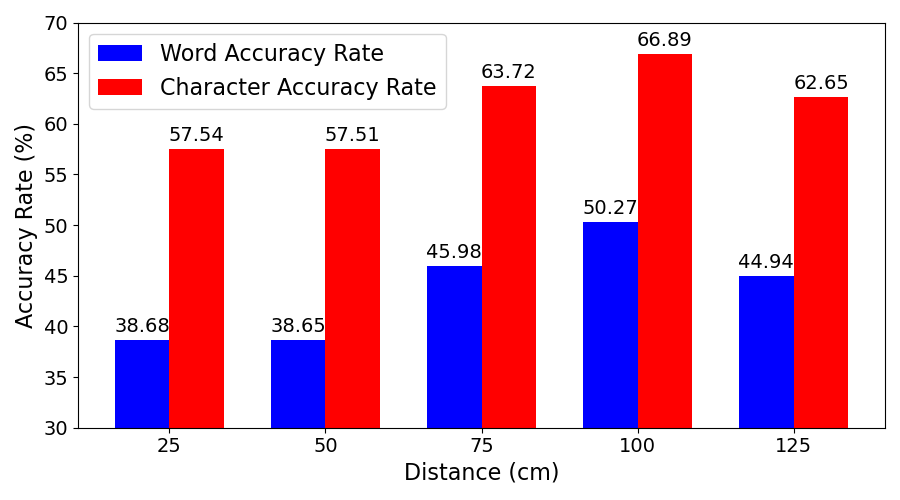}
    \vspace{-0.4cm}
    \caption{Word accuracy rate and character accuracy rate of {\papertitle} at different distances between the radar and smartphone.}
    \label{fig:dist-vs-wacc}
    \vspace{-0.7cm}
\end{figure}

\begin{figure}
    \centering
    \includegraphics[width=0.8\columnwidth]{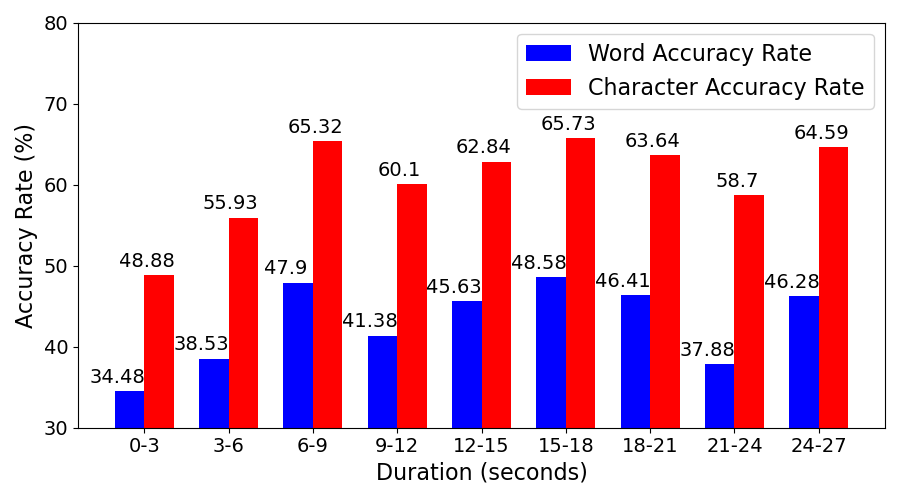}
    \vspace{-0.4cm}
    \caption{Word accuracy rate and character accuracy rate of {\papertitle} for various utterance durations.}
    \label{fig:duration-vs-wacc}
    \vspace{-0.5cm}
\end{figure}

The evaluation of {\papertitle} was conducted from 25 cm to 125 cm, in steps of 25 cm. From each distance, utterances were divided into time segments from 0-3 seconds, 3-6 seconds, and so on, up to 24-27 seconds, to assess how both distance and duration affect word accuracy. A total of 1042 utterances were collected, with data randomly sampled across distances to increase diversity in the test set. The overall word accuracy rate \cite{kolenaWERCER} was \textbf{44.74\%}, and the character accuracy rate was \textbf{62.52\%}. Figures \ref{fig:dist-vs-wacc} and \ref{fig:duration-vs-wacc} show the accuracy rate vs distance, and the accuracy rate vs duration respectively. 

Despite the varying distances and utterance durations, we do not observe a strong correlation between distance or duration and accuracy, which contrasts with the findings from mmSpy \cite{mmspy} and mmEve, where accuracy decreased with increasing distance. We hypothesize that this difference may be due to several factors. First, the LibriSpeech dataset, used in {\papertitle}, features a much greater dynamic range compared to the limited-word datasets like AudioMNIST or Speech Commands used in previous works, causing SNR variations across the data. Additionally, the Whisper-large-v2 model, combined with LoRA, is far more powerful than the models employed in mmSpy or mmEve, likely enabling it to generalize better across longer ranges. Another possible reason is that our system may not be sensitive to the SNR variations within the range of distances we tested, owing to the careful system design, a promising sign that suggests our system performs robustly even in noisy conditions. To gain a complete understanding of the system’s accuracy over longer distances, future experiments will explore ranges up to 500-800 cm. These findings are encouraging and signal the potential for continued improvement through further data collection and validation.

Although the word accuracy rate is lower than conventional ASR systems, it is still valuable in a security context, where even partial word matches pose a significant risk. To demonstrate this, a comparison of transcriptions generated by {\papertitle} and the original LibriSpeech utterances is presented in Table \ref{tab:utterances_table}. As an example, even skilled lip readers from the deaf community can only correctly identify 30-40\% of spoken words \cite{bustleAccurateReading}, yet they maintain basic communication by inferring missing words from context. Similarly, our system's accuracy, comparable to that of lip reading, allows partial word matches to provide enough context to potentially leak sensitive information about the user.

\section{Conclusion and Future Work}\label{sec:conclusion}



This is the first paper that demonstrates mmWave radar's ability to remotely eavesdrop on smartphone conversations, capturing full sentences with an unconstrained vocabulary. In developing {\papertitle}, we addressed challenges like low SNR, limited frequency range, and the lack of large-scale radar-audio datasets through systematic integration of synthetic data generation, domain adaptation of the Whisper-large-v2 model, and advanced signal processing. While promising, this work merely scratches the surface of radar technology’s potential for covert surveillance.

Future work will focus on expanding the system’s word accuracy by exploring context-prompting for ASR models, which could steer predictions toward context-specific words. As radar technology advances, the attack’s effectiveness will likely increase, and similar techniques may also be explored with alternative technologies like lidar. More experiments will be conducted over longer distances (up to 500-800 cm) to fully understand the system's capabilities.

To mitigate these risks, potential defenses could include shielding phone earpieces to minimize detectable vibrations or developing radar signal-jamming techniques to disrupt unauthorized radar-based surveillance. These efforts will be crucial as technology continues to evolve, bringing both new capabilities and security challenges.

\bibliographystyle{IEEEbib}
\bibliography{mmWave-Whisper}

\end{document}